\documentclass[aps,pre,preprint,nofootinbib,groupedaddress,showpacs,showkeys]{revtex4}
\usepackage{epsfig}

\begin{document}
\preprint{APS preprint}
\title{Stochastic Cellular Automata Model for Stock Market Dynamics}

\author{M. Bartolozzi$^{1}$ and A. W. Thomas$^{1}$}
\affiliation{$^1$Special Research Centre for the Subatomic  Structure
of Matter (CSSM), University of Adelaide, Adelaide, SA 5005,
Australia}.  \date{\today}

\begin{abstract}
In the present work  we introduce a stochastic cellular automata model
 in order to simulate the dynamics of the stock market.  A direct
 percolation method is used to create a hierarchy of clusters  of
 active traders on a two dimensional grid.  Active traders are
 characterised by the  decision to buy, $\sigma_{i}(t)=+1$, or sell,
 $\sigma_{i}(t)= -1$, a stock at a certain discrete time step. The
 remaining cells are inactive,  $\sigma_{i}(t)= 0$.  The trading
 dynamics is then determined  by the stochastic interaction between
 traders belonging to the same cluster. Extreme, intermittent events,
 like  crashes or bubbles, are triggered by a phase transition in the
 state of the bigger clusters present on the grid, where almost all
 the active traders come to share the same spin orientation.   Most of
 the stylized aspects of the financial market time series, including
 multifractal proprieties, are reproduced  by the model. A direct
 comparison is made with the daily closures of the S\&P500 index.
\end{abstract}

\maketitle
\section{Introduction}
\label{}

Since the successful application of the Black-Scholes theory for
option pricing~\cite{black73} in 1973 more and more  physicists have
been attracted by the idea of understanding the behaviour of the
market dynamics in terms  of complex system theory, where
self-organized criticality~\cite{soc,bak1997} and stochastic
processes~\cite{mantegna,paul} play important roles. The aim of the
microscopic models proposed so far (for general
reviews~\cite{levy2000,feigenbaum2003}) is to reproduce some  {\em
stylized facts}~\cite{sornette99} concerning the temporal fluctuations
of the price indices, $P(t)$. In particular, the logarithmic price
returns

\begin{equation}
R(t)=ln P(t+1)- ln P(t),
\label{returns}
\end{equation}
and the volatility, defined in the present work as
\begin{equation}
 v(t)=|R(t)|,
\end{equation}
have been studied extensively~\cite{mantegna} from an empirical point
of view.  The results have shown that while long time correlations are
present in the volatility, a phenomenon  known as {\em volatility
clustering}, they cannot be found in the  time series of
returns. Moreover the latter show  an intermittent behaviour that
recalls in some aspects hydrodynamic
turbulence~\cite{mantegna97,mantegna95,frisch},  characterized by
power law tails in the probability distribution function (pdf).
Microsimulations have demonstrated that this kind of behaviour can
originate both as a  stochastic process with multiplicative
noise~\cite{kaizoji,krawiecki02,takayasu9798} and as a  percolation
phenomenon~\cite{stauffer,stauffer98,cont00}.

In order to reproduce these features of real markets  we introduce a
stochastic cellular automata model, representing an {\em open}
market. That is, a market where the number of {\em active} traders,
defined as cells with spin state different from 0, namely
$\sigma_{i}(t)= \pm 1$, evolve in time according to a percolation
dynamics.  The percolation dynamics is chosen in order to simulate the
herding behaviour typical of investors~\cite{sornette03}.  According
to this, active traders gather in {\em clusters} or {\em networks}
where, following a stochastic exchange of information,  they formulate
the trading strategy for the next time step. The  results obtained by
the simulations are then compared with the time series  of daily
closures of the S\&P500  index
~\cite{mantegna,plerou99,cizeau97,gopikrishnan99,liu99,sornette03,ausloos2003,Tsallis2003,canessa2000}
over a period of about 50 years.

Moreover, recently, the fractal properties~\cite{feder} of the  price
fluctuations have also been   investigated for different
markets~\cite{rodrigues01,gorski02,auloos02,dimatteo03}.  A common
feature found in these studies is the existence of a nonlinear,
multifractal spectrum that excludes the possibility  of {\em efficient
market} behaviour~\cite{bachelier00}.  The origin of the
multifractality of in the financial time series has also been  at the
center  of
discussions~\cite{lux2001,mandelbrot2001,lebaron2001,stanley2001}. In
this paper  we consider the multifractal spectrum of the price
fluctuations as a stylized fact of the market time series without
addressing any question about the underlying process able to generate
it. The multifractal spectrum is used as a further test for our model.

A parallel between multifractal and thermodynamical formalism has also
been investigated.  We found, in agreement with the previous work of
Canessa~\cite{canessa2000}, that the {\em analogue specific heat} can
provide a good tool to characterize intermittency, that is financial
crashes or bubbles, from a thermodynamics-equivalent point of view.

\section{The Model}
In the present work we simulate the financial market dynamics  via a
stochastic cellular automata model.  The agents of the market  are
represented by  cells  on a two dimensional grid, 512x128.  The  $i$th
agent at the discrete time step $t$ is characterized by three possible
states or  spin orientations, $\sigma_{i}(t)=0,\pm 1$.  The value
$\sigma_{i}(t)=+1$ is associated with the purchase of a stock while
$\sigma_{i}(t)=-1$ with selling. The former  states are called {\em
active}.  The cells with spin value $\sigma_{i}(t)=0$ are {\em
inactive} traders.  The active traders herd in {\em networks} or {\em
clusters} via  a direct  percolation method related to a {\em forest
fire} model~\cite{stauffer}.  The information carried by the active
traders,  that is their spin state, is shared  with the other members
of the cluster.  The percolation dynamics allow a time dependent
herding behaviour and the market can be interpreted as an open system
not bounded by conservation laws.  The clustering process will be
discussed  in detail in the next subsection.
 
The trading dynamics is instead related to the synchronous update of
the  spins of the active traders,  ruled by a stochastic exchange of
information between them, similar to a random Ising
model~\cite{kaizoji,krawiecki02}.  A particular feature of the present
simulation is that the information is not spread all over the grid,
as in other multi-agent scenarios~\cite{kaizoji,krawiecki02}, but  it
is limited by the clusters of interaction previously defined.  The
mechanism for the spin dynamics is explained in  ``stochastic trading
dynamics'' subsection.

\subsection{Percolation Clustering}

One of the aims of our cellular automata model is to reproduce the
{\em herding behaviour} of active traders~\cite{sornette03}.  We refer
to herding behaviour as the tendency of people involved in the market
to aggregate in networks or clusters of influence.  The traders then
use the information obtained by their network in order to formulate a
market strategy.  Even if the topological structure of these networks
of information  is not important, since several kinds of long range
interaction  are available nowadays~\cite{krawiecki02}, the number of
connections for each trader must be, in any case, finite and not
extended over the whole market.  In this framework a direct
percolation method is used to simulate  herding dynamics between
active traders. If we assume that the neighbours of  influence are
those of von Neumann (up, down , left, right), the percolation is
fixed by the following parameters:

{\bf $p_{h}$}: the probability that an active trader can turn one of
his inactive neighbours into an active one at  the next time step,
$\sigma_{i}(t)=0 \rightarrow \sigma_{i}(t+1)= \pm 1 $.  This simulates
the fact that certain information possessed by a trader    may induce
a {\em potential} trader to join the market dynamics.

{\bf $p_{d}$}: the probability that an active trader {\em diffuses}
and so becomes inactive, $\sigma_{i}(t)= \pm 1 \rightarrow
\sigma_{i}(t+1)=  0 $, due to each of his inactive neighbour.
This mimics the fact that only traders at  the borders of a network,
that is the weaker links, can quit the market.

{\bf $p_{e}$}: the probability that a non trading cell spontaneously
decides to enter  the market dynamics, $\sigma_{i}(t)=0 \rightarrow
\sigma_{i}(t+1)= \pm 1 $.

\begin{figure}
\centerline{\epsfig{figure= 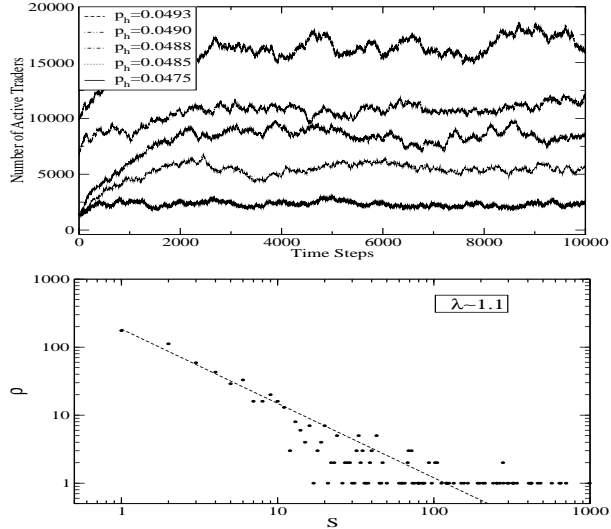,height=7cm, width=8cm}}
\caption{ Top: Different values of the parameter {\em H} produce
different activities of traders on the grid.  Bottom: Cluster size
distribution for $p_{h}=0.0485$ at $t=9000$.}
\label{fig1}
\end{figure}

The values of the adimensional parameters, $p_{h}$, $p_{d}$ and
$p_{e}$, influence  the stability of the system and the percentage of
active traders on the grid.  In order to test different market
activities  we fix the values $p_{d}=0.05$ and $p_{e}=0.0001$ while we
tune the parameter $p_{h}$.  At the beginning of the simulation the
grid is loaded randomly with a small percentage of active traders and
then the system is permitted to evolve according to the previous
rules.  If we are in a stable range of the parameter $p_{h}$, after a
transient period  that depends both on the parameter values and the
initial  number of active cells, the number of active traders on the
grid  begins to fluctuate around a certain average, as shown in
Fig. \ref{fig1} (Top).  The market can be considered  {\em  open}
since the number of agents changes  dynamically in time. In this
regime, the competition between herding and diffusion produces a power
law distribution  of the cluster size, as shown in Fig.~\ref{fig1}
(Bottom), $\rho (S) \approx S^{-\lambda}$, where $S$ is the cluster
dimension, defined as the number of active cells belonging to the same
cluster, and $\lambda>0$, creating a  hierarchy of networks.  This
hierarchy is necessary if were are to take into account  a real aspect
of the market, namely  that different traders also have different
trading powers.  A reasonable assumption is that people having a
larger number of sources of information, so belonging to greater
clusters, can be associated  with professional investors that, most
likely, are able to move a greater amount of stocks compared to the
occasional investor. Using this assumption  we are able to define a
proper weight for the trading power of different cells, as we will
discuss in the next subsection.

 A similar percolation model has also been used to reproduce some
statistical and geometrical features of solar
activity~\cite{wentzel,vlahos}.

\subsection{Stochastic Trading Dynamics}

The dynamics of the spins of the active traders, $ \sigma^{k}_{i}(t)=
\pm 1$  for $i=1,..., N^{k}(t)$ (where the superscript $k$, from now
on, refers  to the $k$th cluster of the grid configuration at time
step $t$) follows a stochastic process that  mimics the human
uncertainty in decision making~\cite{krawiecki02}.  Their values are
updated synchronously according to a local probabilistic rule:
$\sigma^{k}_{i}(t+1)=+1$ with probability $p^{k}_{i}$ and
$\sigma^{k}_{i}(t+1)=-1$  with probability $1-p^{k}_{i}$.  The
probability $p^{k}_{i}$ is determined, by analogy to heat bath
dynamics with formal temperature $k_{b}T=1$, by
\begin{equation}
 p^{k}_{i}(t)=\frac{1}{1+e^{-2I^{k}_{i}(t)}},
\end{equation} 
where the local field, $I^{k}_{i}(t)$, is
\begin{equation}
I^{k}_{i}(t)=\frac{1}{N^{k}(t)}\sum_{j=1}^{N^{k}(t)}A^{k}_{ij}\sigma^{k}_{j}(t)
+h^{k}_{i}.
\end{equation}  
The $A^{k}_{ij}(t)$ are time dependent interaction strengths between
agents  and $h^{k}_{i}(t)$ is an external field reflecting the effect
of the  environment~\cite{krawiecki02}.  The interaction strengths and
the external field change randomly in time according to
$A^{k}_{ij}(t)=A \xi^{k}(t) +a \eta_{ij}(t)$ and
$h^{k}_{i}(t)=h\zeta^{k}_{i}(t)$.  The variables
$\xi^{k}(t),\eta_{ij}(t),\zeta^{k}_{i}(t)$ are random  variables
uniformly  distributed in the interval (-1,1) with no correlation in
time or space.  The measure of the strengths of the previous terms,
$A$, $a$ and $h$, are  constant and common for all the grid.
 
In this contest the dynamics of the price index, $P(t)$, can be easily
derived if we assume that the index variation is proportional to the
difference between demand and supply,
\begin{equation}
 \frac{dP}{dt} \propto xP,
\label{price}
\end{equation}
and using a weighted average for the orientation of the spins,
\begin{equation}
x(t)= \beta \sum_{k=1}^{N_{cl}(t)}\sum_{i=1}^{N^{k}(t)} N^{k}(t)
\sigma^{k}_{i}(t),
\label{orientation}
\end{equation}
where $N_{cl}(t)$ and $N^{k}(t)$ are, respectively,  the number of
clusters on the grid and the size of the $k$th cluster,  while $\beta$
is a normalization constant. The relation (\ref{orientation}) follows
from the assumption  of proportionality between the financial power of
an active cell and the size  of the cluster to which it belongs. A
justification of this relation  is that professional investors, able
to move larger amounts of capital,  are more likely to be linked with
a large  number of other investors than the occasional traders.  From
(\ref{price}) we find that the logarithm  of the price returns
(\ref{returns}), that is the  fundamental quantity we aim to model, is
proportional to the average orientation  defined in
(\ref{orientation}), $R(t) \propto x(t)$.

\section{Numerical Results}

We compare the results of our model with the  Standard \& Poor 500
(S\&P500) index that is one of the most widely used benchmarks for US
equity performance.  The database analyzed is composed by of the daily
indices  from 3/1/1950 to 18/7/2003 for a total of 13468 data.  The
time series of the index prices, $P(t)$, is  converted into  the
logarithmic returns (\ref{returns})  and then  normalized over the
time interval, $T$,
\begin{equation}
r(t)=\frac{R(t)-\langle R(t)\rangle_{T}}{\sigma (R(t))}.
\end{equation}
where $\langle \ldots \rangle_{T} $ denotes the temporal average and
$\sigma$ is the standard deviation. In this way all the sets of
returns  will have zero average and $\sigma=1$.

Before discussing the results we briefly describe the behaviour  of
the cellular automata with respect to changes of the parameters.
Regarding the percolation, the herding parameter $p_{h}$, as
previously seen,  determines the concentration of active traders on
the grid. For a low concentration the active traders will be
distributed in small clusters and the information  will be extremely
split up  on the grid. In this case large coherent events will be more
rare.  A higher concentration, where clusters of the order of  a
thousand cells are present, allows large events to occur with higher
frequency. In fact, the phase transition of large clusters can easily
trigger a  crash or a bubble in the market.  Time series related to a
higher herding parameter  are therefore more intermittent.  We can
then infer that a market made by small groups of traders behaves like
a {\em  noisy market}, while big crashes or bubbles must necessarily
be related to the  interaction of large networks and so to a kind of
crowd behaviour. For most of the simulations  we fixed the parameter
$p_{h}=0.0493$ in order to have bigger clusters, with $S^{max}(t)$  of
the order of 2500 cells.
%
\begin{figure}
\centerline{\epsfig{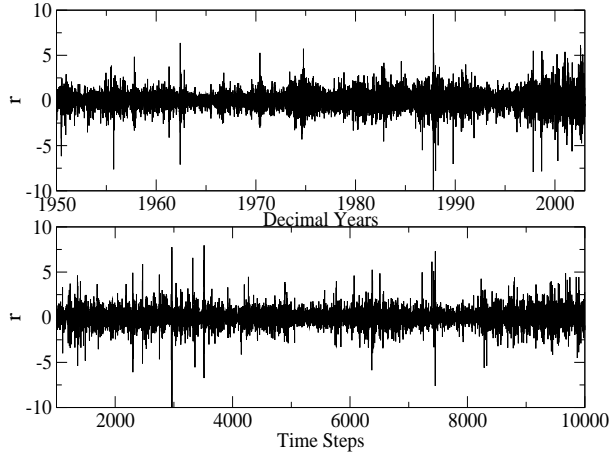}}
\caption{ Top: Normalized logarithmic returns for the S\&P500.
Bottom: Time series of returns reproduced with the simulation with
$p_{h}=0.0493$, $A=1.8$ and $h=0$.}
\label{fig2}
\end{figure}

The strength $A$ also plays an important role in the trading dynamics.
With $p_{h}$ fixed, this parameter is related to the  intermittency of
the system.  Both for large values of the activity ($A > 10$) and for
$A \rightarrow 0$ we observe  an approach of the pdf toward a
Gaussian-like shape. That is very, large fluctuations become more and
more rare, and  $A$ can be regarded as a temporal scale for the
system, similar  to the activity parameter in the Cont-Bouchaud
model~\cite{cont00}.  In spite of this some large fluctuations can be
still identified.  This is probably one of the main  differences
between the Cont-Bouchaud model and the present.  In fact Monte Carlo
simulations of  the former~\cite{stauffer98} show that an increase of
the activity  brings a rapid convergence toward a Gaussian
distribution because a large number of clusters are trading at the
same time following a random procedure of decision
making~\cite{cont00, stauffer98}: there are no clusters that can
influence the  market more than others and so the resulting global
interaction is noise-like.  In our model fluctuations are always
allowed because of the   heat bath dynamics. Clusters of active
traders can always be subjected to phase transitions, independently
of the state of other clusters, creating a displacement between demand
and supply.

In order to reproduce the behaviour of financial time series we work
with $1.5<A<2.5$.  In this range there is a nonlinear dependence of
the intermittency on $A$. The value of the parameter $a$ is fixed by
the relation $a=2A$, already used in the work of Krawiecki et
al.~\cite{krawiecki02}. However, we note that the ratio $a/A$ is
actually not essential in reproducing the intermittency found in real
data.  Even the parameter $h$ does not have a central role  in the
simulation, as long as $h<<1$.  In fact, the percolation dynamics of
the active traders already  introduces a  natural noise. In most of
the runs we simply set $h=0$.
 
\begin{figure}
\centerline{\epsfig{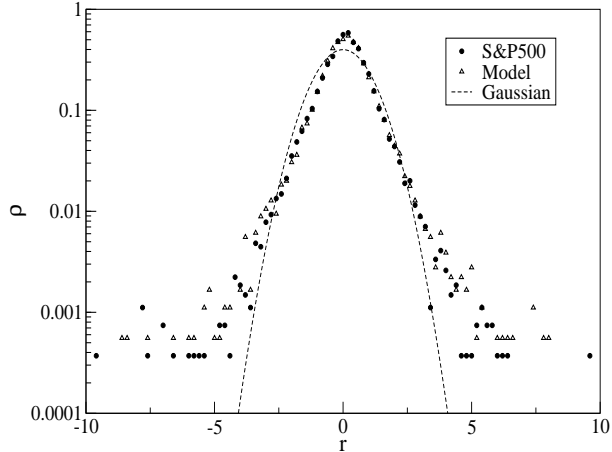}}
\caption{ Probability distribution function for the S\&P500 computed
with daily data from 31/1/1950 to 18/7/2003 and the model. A Gaussian
is also plotted for comparison. The parameters used are
$p_{h}=0.0493$, $A=1.8$ and $h=0$.}
\label{fig3}
\end{figure}

Now we discuss the results of  the cellular automata.  In
Fig. \ref{fig2} (Top) and Fig. \ref{fig2} (Bottom)  the normalized
logarithmic returns of the S\&P500 and of the simulation are shown,
respectively. The average number of active traders, in the stable
regime, is $\approx 16000$, as shown in Fig. \ref{fig1} (Top). The
model reproduces  the intermittent behaviour of the S\&P500 time
series, as expressed  by the {\em leptokurtic} pdf of Fig.\ref{fig3}.
The tails of the distribution follow a power law decay, reflecting the
fact that large coherent events, far from the average,  are likely to
occur with a frequency higher than expected for a random process
(where the shape would be a Gaussian). These large events are related
to financial crashes or bubbles of the market and, in our model, to a
phase transition in the spin state of large networks of active
traders, as we will discuss further on.  From a power law fit, $\rho
(r) \approx r^{-1-\gamma}$   (for $|r|>2$), we find $\gamma \approx 3$
for both the S\&P500 and the model, confirming the  good agreement
between the two.

The problem of finding the best distribution describing the price
returns is a very important issue from a practical point of
view~\cite{mantegna}.  The standard Black-Scholes theory for option
pricing~\cite{black73,mantegna,paul} assumes that the returns are
normally distributed. This fact has been proven to be empirically
false, as shown also in Fig. \ref{fig3} (see Ref.~\cite{mantegna} for
a general reference).  Finding a more appropriate distribution would
be an important  improvement in this field of research\footnote{An
intriguing framework of investigation has been provided by the {\em
non-extensive statistical mechanics} proposed by
Tsallis~\cite{Tsallis1988,Tsallis1999,tsallis_market,ausloos2003,Tsallis2003}.
A more complete discussion on this important topic is beyond the scope
of this paper and will be discussed in  future work.}.

%

\begin{figure}
\centerline{\epsfig{figure= 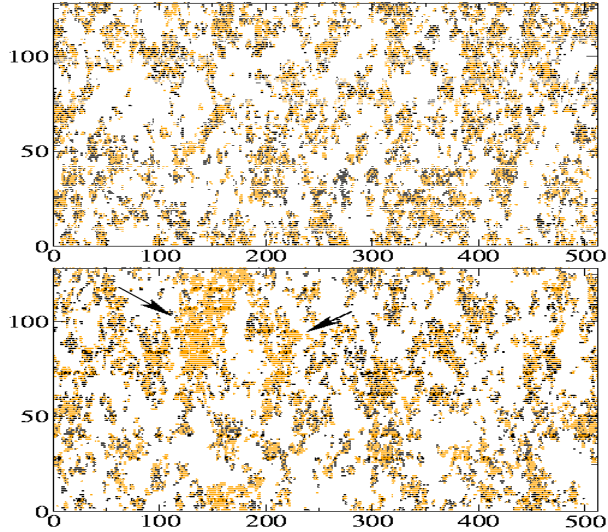,height=7cm, width=8cm}}
\caption{ Top: snapshot of the grid during a normal trading
period. The  black cells are the buyers while the orange ones are the
sellers. The parameters used in this simulation are $p_{h}=0.0493$,
$A=1.8$ and $h=0$.  Bottom: the same simulation during a crash. Large
clusters of sellers are indicated by arrows.}
\label{fig4}
\end{figure}

In order to understand the trading dynamics of  the automata we also
show  two snapshots of the grid configuration, the first during a
normal session, Fig.\ref{fig4} (Top), and the second  during a crash,
Fig.\ref{fig4} (Bottom).  During the normal session the orientations
of  the spins are distributed uniformly over the various clusters and
there is no sharp  difference between demand and supply. The situation
is different during a crash. In this case the clusters at the top of
the hierarchy, the bigger ones, play a fundamental role. In fact they
undergo  a phase transition where the greatest part of  their spins
share the same orientation.  The capacity of the clusters to generate
a coherent orientation of the spins, and hence of their trading state,
can be interpreted in terms  of a multiplicative noise
process~\cite{paul,liu99}, where the collective synchronization arises
as a result of the randomly varying interaction strengths between
agents.  The peculiarity of our model is that crashes or bubbles
(sudden price changes) are related not to a phase transition of the
whole market~\cite{krawiecki02,kaizoji}  but rather to a phase
transition in one or more of the larger clusters that have a greater
influence on the trading session. This behaviour is probably closer to
the real market where the synchronization of trading opinion is more
likely to happen between large groups of traders than over the whole
market.
   
%

\begin{figure}
\centerline{\epsfig{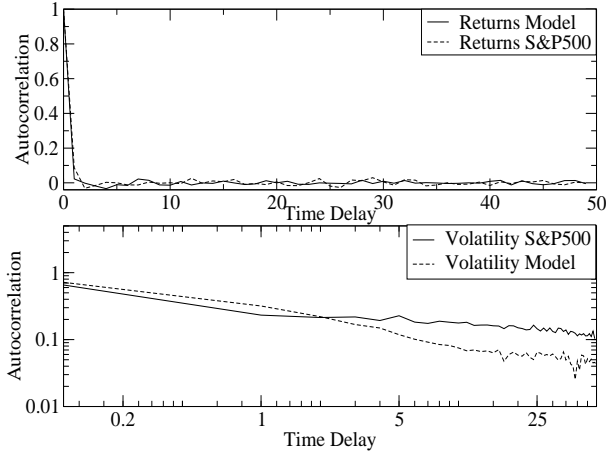}}
\caption{ Top: Autocorrelation function for the price returns. Bottom:
 Autocorrelation function for the volatilities.  In both the graphs
 the parameters used for the model are $p_{h}=0.0493$, $A=1.8$ and
 $h=0$.}
\label{fig5}
\end{figure}

The temporal correlations of the logarithmic returns and of the
volatility  are investigated via the autocorrelation function, defined
as
\begin{equation}
c(\tau)=\sum_{t=1}^{T-\tau}x(t+\tau)x(t),
\end{equation}
where {\em T} is the length of the time series and $\tau$ is a time
delay for the normalized variable $x(t)$. The results for both the
model and the S\&P500 are shown in Fig.\ref{fig5} (Top) and
Fig.\ref{fig5} (Bottom), respectively.  While the temporal
correlation for the returns is lost almost immediately, the volatility
manifests a slow decay in time, related to the phenomenon of
volatility clustering.  The previous temporal dependencies have been
found in both real data and in the simulation.
%

\section{Multifractal Analysis}
It is also worth pointing out that financial time series present an
inherent {\em multifractality}~\cite{feder}.  In the past few years
the work of many
authors~\cite{rodrigues01,gorski02,auloos02,dimatteo03} has been
addressed to the characterization of the multifractal properties of
financial time series, and nowadays multifractality can be considered
as a stylized fact.  In order to study the multifractal properties of
our model we use the {\em generalized Hurst
exponent}~\cite{mandelbrot}, $H(q)$, derived via the $q-$order
structure function,
\begin{equation}
S_{q}(\tau)= \langle |x(t+\tau)-x(t)|^{q}{\rangle_{T}} \propto
\tau^{qH(q)},
\label{structure}
\end{equation} 
where $x(t)$ is a stochastic variable over a  time interval $T$ and
$\tau$ the time delay.  The generalized Hurst exponent, defined in
(\ref{structure}),  is an extension of the Hurst exponent, $H$,
introduced in the context of reservoir control on the Nile river dam
project,  around 1907~\cite{feder,hurst51}. This technique   provides
a sensitive method for revealing long-term  correlations in random
processes.  If $H(q)=H$ for every $q$ the process is said to be
monofractal and $H$ is equivalent to the original definition of the
Hurst exponent. This is the  case of simple Brownian motion or
fractional Brownian motion.

\begin{figure}
\centerline{\epsfig{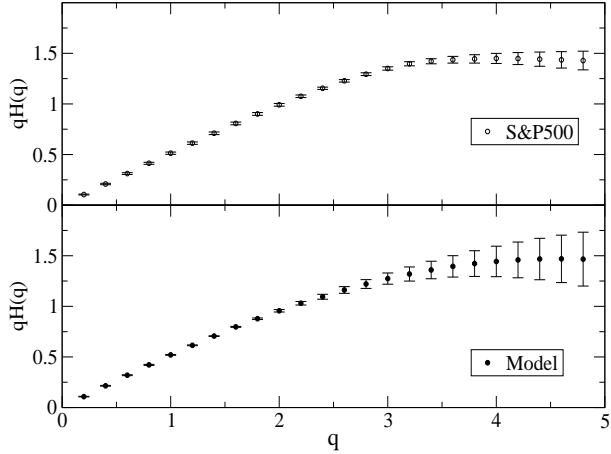}}
\caption{ Multifractal spectra for the S\&P500 in the period
31/1/1950 to 18/7/2003 (top) and the model (bottom).  For the latter
$p_{h}=0.0493$, $A=1.8$ and $h=0$.}
\label{fig6}
\end{figure}

If the spectrum of $H(q)$ is not constant with $q$  the process is
said to be multifractal.  From the definition (\ref{structure}) it is
easy to see that  the function $H(1)$ is related to the scaling
properties of the volatility.  By analogy with the classical Hurst
analysis, a phenomenon is said to be persistent if $H(1)>1/2$ and
antipersistent if $H(1)<1/2$.  For uncorrelated increments, as in
Brownian motion, $H(1)=1/2$.  In Fig.~\ref{fig6} a comparison is shown
between the multifractal spectra of the  model and the S\&P500
obtained from the prices time series.  It is clear that both processes
have a multifractal structure and the price fluctuations cannot be
associated  with a simple random walk as in the classical  {\em
efficient market hypothesis}~\cite{bachelier00}.
%

The multifractality of the time series can also be discussed  in terms
of thermodynamic equivalents, according to multifractal
physics~\cite{lee88,canessa1993,vollhardt97,canessa2000}. In this
approach we divide the time series, $x(t)$ , for $t=1,...,L$ into
$\mathcal{N}$ equal sub-intervals.  Then we can write the following
measure for each of these,

\begin{equation}
\mu_{i}(\tau)=\frac{|x(t+\tau)-x(t)|}{\sum_{t=1}^{\mathcal{N}}|x(t+\tau)-x(t)|},
\end{equation}
with $i=1,...,\mathcal{N}$ and $\tau$ the time delay.  The quantity
 $\mu_{i}(\tau)$ can be viewed as a normalized  probability
 measure. The corresponding generating function is given by

\begin{equation}
Z(q,\mathcal{N})=\sum_{i=1}^{\mathcal{N}}\mu_{i}(\tau)^{q} \propto
\mathcal{N}^{-\chi_{q}},
\end{equation}
that is an analogue of the partition function in thermodynamics.
According to Ref.~\cite{lee88} the scaling exponent $\chi_{q}$ is
directly related  to the generalized multifractal dimension
$D_{q}$~\cite{feder},

\begin{equation}
\chi_{q} \equiv (q-1)D_{q}.
\label{chi}
\end{equation}

If we consider $\chi_{q}$ as the {\em free energy} of our system,
Eq.(\ref{chi}) provides a link between the classical thermodynamical
formalism and multifractality.  Assuming $q$ as an equivalent
temperature, we can define an {\em analogue specific
heat}~\cite{lee88,canessa1993,canessa2000,ivanova02},

\begin{equation}
C_{q} = -\frac{\partial^{2}\chi_{q}}{\partial q^{2}}.
\label{spec_heat}
\end{equation}

The previous equation for $C_{q}$ can also be written in terms of
singular measure formalism~\cite{davis94}. In this framework we define
a measure, $\epsilon(1,t)$, as

\begin{equation}
\epsilon(1,t)=\frac{|x(t+1)-x(t)|}{\frac{1}{L-1}\sum_{t=1}^{L-1}
|x(t+1)-x(t)|}.
\end{equation}

\begin{figure}
\centerline{\epsfig{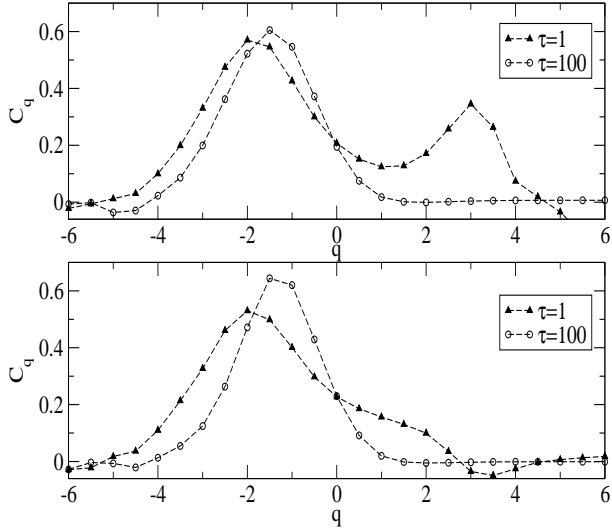}}
\caption{ Top: analogue specific heat for the S\&P500 from 31/1/1950
to 18/7/2003 for two different time delays, namely $\tau=1$ and
$\tau=100$. A sharp peak is clearly visible around $q=-1.5$. The
second peak on the right hand side disappears  increasing the temporal
delay. The $C_{q}$ curves have been computed for the logarithm of the
price using  the algorithm in~\cite{davis94} for $K_{q}$.  Bottom:
analogue specific heat for the model with parameters $p_{h}=0.0493$,
$A=1.8$ and $h=0$. The double humped shaped for small temporal delays
is visible also here.}
\label{fig7}
\end{figure}

We can then generate a series of measures on shorter intervals of
length $\delta$, $\epsilon(\delta,l)$ where $\delta$ is an integer
power of two and $l$ is the index  of the first point of the sub
segments at that resolution.  The average measure in the interval
[$l,l+\delta$] is

\begin{equation}
\epsilon(\delta,l)=\frac{1}{\delta}\sum_{l^{\star}=l}^{l+\delta-1}
\epsilon(1,l^{\star}),
\end{equation}
for $l=0,...,L-\delta$.  In this case we have a scaling property for
 the ensemble average with  respect to the scale $\delta$:

\begin{equation}
\langle \epsilon (\delta,l) \rangle \propto \delta^{-K_{q}}.
\end{equation}
In a multifractal process the exponent $K_{q}$ is a nonlinear function
of $q$ related to the intermittency of the time series and to the
generalized dimension via~\cite{hentschel83,halsey86}

\begin{equation}
(q-1)D_{q}=q-1-K_{q}.
\label{kq}
\end{equation} 
From (\ref{spec_heat}) and (\ref{kq}) we have that~\cite{ivanova02},

\begin{equation}
C_{q} = \frac{\partial^{2}K_{q}}{\partial q^{2}}.
\end{equation}
Following Ref.~\cite{canessa2000} we have found the analogue specific
heat for both the S\&P500 and for the price time series generated by
our model, see Fig.\ref{fig7}.  For $\tau=1$ we observe a
double-humped shape  for  both the model and the empirical data. If we
take longer time delays the shoulder on the right hand side
disappears, leaving only a sharp peak, similar to a first order phase
transition around $q=-1.5$. The   results are in agreement with the
analysis of Canessa~\cite{canessa2000} and recall the Hubbard model
for small to intermediate values of the  local
interaction~\cite{vollhardt97}.   As also suggested
in~\cite{canessa2000}, the second peak is due to the large
fluctuations at small scales, that is crashes and bubbles. Increasing
the time delay  means that the fluctuations tend to be smoothed and
the time series of returns approach  a noise-like regime. For this
reason the analogue specific heat shapes for $\tau=100$ are basically
indistinguishable.  From this argument we can interpret the analogue
specific heat, and in particular the second peak for low time delays,
as a way to characterize crashes, or in general the degree of
intermittency in a time series. The difference in the shapes and
heights of the shorter peak for $\tau=1$ is due to a slightly
different correlation  of the fluctuations in the two time series.
Moreover, in the model we can link the shorter peak to the physical
phase transition in the spin state of a network of traders.

\section{Conclusions}

In this paper we have introduced a stochastic cellular automata model
for the dynamics of the financial markets.  The main difference
between our model and other stochastic simulations based on spin
orientation of agents~\cite{krawiecki02,kaizoji} is the temporal
evolution of the networks of interaction and therefore the concept  of
an ``open'' market.  The active traders follow a direct percolation
dynamics  in order to aggregate in networks of information.  This
makes our simulation, even if still a raw approximation,  surely
closer to the real market, where no conservation rules for the number
of agents can be claimed. Crashes and bubbles can be interpreted  as a
synchronization of the spin orientation of the more influential
networks in the market.  Moreover, the introduction of a limitation in
the number of interacting agents reduces drastically the number of
computations on the grid per time step. In a system where all  the
agents interact with each other this number goes  like $N_{a}^{2}$,
being $N_{a}$ the number of active agents, while in our model,
considering the distribution of the clusters, it is easy to see that
it goes like $N_{a}^{2-\lambda}$. The value of $\lambda$ found for
several herding parameters, $p_{h}$,   is $\lambda \approx 0.6 \div
1.4$, so that the computational cost is much lower for the present
model.  This gives one the possibility to simulate the market using a
very large range of agents.  The model is able to reproduce most of
the stylized  aspects of the financial time series, supporting the
idea that crashes and bubbles are related to a collective
synchronization in the trading  behaviour of large networks of
traders,  where the information is exchanged according  stochastic
interaction between them.

\begin{acknowledgments}
This work was supported by the Australian Research Council.
\end{acknowledgments}

\end{document}